\documentclass[english,prl,aps,reprint,longbibliography]{revtex4-1}
\usepackage[T1]{fontenc}
\usepackage[latin9]{inputenc}
\usepackage{amsbsy}
\usepackage{amstext}
\usepackage{graphicx}
\usepackage{esint}

\makeatletter

\newcommand*\LyXThinSpace{\,\hspace{0pt}}

\makeatother

\usepackage{babel}
\begin{document}
\title{Half-Quantized Hall Effect at the Parity-Invariant Fermi Surface}
\author{Jin-Yu Zou, Rui Chen}
\affiliation{Department of Physics, The University of Hong Kong, Pokfulam Road, Hong Kong, China}
\author{Bo Fu}
\email{fubo@hku.hk}

\affiliation{Department of Physics, The University of Hong Kong, Pokfulam Road, Hong Kong, China}
\author{Huan-Wen Wang, Zi-Ang Hu}
\affiliation{Department of Physics, The University of Hong Kong, Pokfulam Road, Hong Kong, China}
\author{Shun-Qing Shen}
\email{sshen@hku.hk}

\affiliation{Department of Physics, The University of Hong Kong, Pokfulam Road, Hong Kong, China}
\date{\today}
\begin{abstract}
Condensed matter realization of a single Dirac cone of fermions in two dimensions
is a long-standing issue. Here we report the discovery of a single gapless Dirac
cone of half-quantized Hall conductance in a magnetically-doped topological insulator
heterostructure. It demonstrates that the Hall conductance is half-quantized in the
unit $e^{2}/h$ when the parity symmetry is invariant near the Fermi surface. The
gapless Dirac point is stable and protected by the local parity symmetry and the
topologically nontrivial band structure of the topological insulator. The one-half
Hall conductance observed in a recent experiment {[}Mogi et al, Nat. Phys. 18, 390
(2022){]} is attributed to the existence of the gapless Dirac cone. The results suggest
a condensed matter realization of a topological phase with a one-half topological
invariant.
\end{abstract}
\maketitle

\paragraph*{Introduction}

Search for a single gapless Dirac cone of fermions is a long standing issue in condensed
matter physics \citep{Murakami-07njp,Yang-14nc,Armitage-18rmp}. In quantum field
theory, an ideal massless two-dimensional Dirac fermion coupled to a U(1) gauge field
gives rise to the parity anomaly, characterized by a half-quantized Hall conductance
\citep{Niemi1983axial,Jackiw1984Fractiona,Redlich1984gauge,Boyanovsky1986physical,Schakel1991relativistic}.
Lattice regularization of a single gapless Dirac cone on a lattice is not realizable
if the parity symmetry is invariant according to the fermion doubling theorem \citep{Nielsen-81}.
One possible scheme is Wilson fermions which possess linear dispersion near the energy
crossing point, but break the time reversal symmetry at higher energy \citep{Wilson-75,Rothe}.
The proposal for the condensed matter realization of parity anomaly dates back to
1980s \citep{SemenoffPRL1984,FradkinPRL1986,Haldane1988Model}. In his seminal paper
\citep{Haldane1988Model}, Haldane proposed that when one of the two valleys on a
honeycomb lattice is finely tuned to be closed while another one remains open, a
single flavor of massless Dirac fermion with parity anomaly can be realized. It is
actually a critical transition point between a quantum anomalous Hall insulator and
a conventional insulator. In graphene, the parity anomaly with half-integer quantum
Hall effect is masked in view of the fourfold degeneracy from spin and valley in
the system \citep{Neto-09rmp}. Due to the presence of the parity symmetry, the paired
Dirac cones give rise to contributions to the anomaly terms with opposite signs thus
exhibit no anomaly as a whole although the integer quantum Hall effect in graphene
in a magnetic field has some clue of parity anomaly \citep{Novoselov-05nature,Zhang-05nature}.
A three-dimensional topological insulator hosts a single Dirac cone of fermions on
its surface \citep{Fu2007topological,hasan2010colloquium,qi2011titsc,shen2012topological}.
It provides a possible platform to observe the half-quantization of the Hall conductance,
and a lot of attempts have been made in the direction \citep{Fu-07prb,Qi2008topological,Essin-09prl,Chu-11prb,Koenig2014half,XuY-14np,Zhang2017anomalous,B=0000F6ttcher2019survival}.
Recently, the observation of the half-quantized Hall conductance in transport at
zero magnetic field was reported as a signature of the parity anomaly in a semi-magnetic
topological insulator heterostructure \citep{Mogi-21np}. The paired gapless Dirac
cones in a topological insulator thin film are located separately on the top and
bottom surfaces. The local time reversal symmetry breaking on one surface by magnetic
doping may open an energy gap for the Dirac surface fermions while the Dirac fermions
remain gapless on the other surface. Existing theories suggest that the massive Dirac
fermions give rise to half-quantized Hall conductance \citep{Fu-07prb,Qi2008topological}.
However, it is known that all the independent bands on a two-dimensional finite Brillouin
zone just have an integer Chern number \citep{TKNN,Xiao-10rmp}. The gapless Wilson
fermions in two-dimensions have a half-quantized Hall conductance when the valence
bands are fully filled \citep{FuB-22xxx,Zou-22prb}. Thus the semi-magnetic topological
insulator becomes a potential candidate to realize a single gapless Dirac cone in
condensed matter.

\begin{figure}
\includegraphics[width=8.5cm]{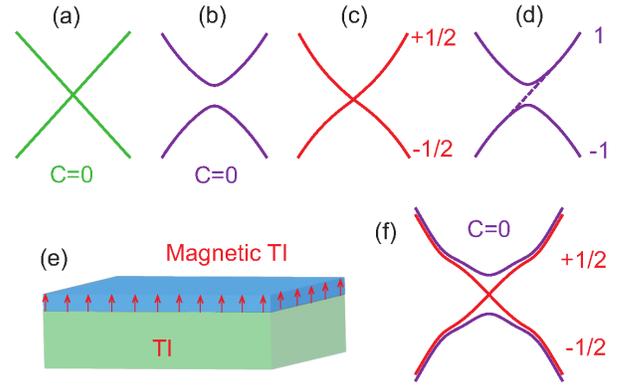} \caption{Four types of two-dimensional Dirac fermions and parity anomalous semimetal in a
semi-magnetic topological insulator. (a) An ideal Dirac fermions with the Hall conductance
$C=0$ in the unit of $\frac{e^{2}}{h}$, which cannot be realized on a lattice according
to the fermion doubling theorem \citep{Nielsen-81}. (b) Topologically trivial gapped
Dirac fermion with $C=0$; (c) The gapless Dirac fermions of linear dispersion at
the Dirac point with $C=\pm1/2$; (d) Topologically nontrivial gapped Dirac fermion
with $C=\pm1$, i.e., Chern insulator. (e) Schematic of a semi-magnetic topological
insulator film. (f) Schematic of the band structure in a semi-magnetic topological
insulator, \textit{i.e.}, parity anomalous semimetal.}
\end{figure}

In this Letter, we report the discovery of the gapless Dirac cone of half-quantized
Hall conductance in a semi-magnetic topological insulator heterostructure. The main
results are summarized in Fig.1. The gapless Dirac cone always has one half quantized
Hall conductance in the units of $e^{2}/h$ when the parity symmetry is invariant
near the Fermi surface. The gapless Dirac point is protected by the local parity
symmetry and the topologically nontrivial band structure of the topological insulator,
although the parity symmetry was broken at higher energy. The gapped Dirac cone has
a nonzero Hall conductance, but becomes zero when the band is fully filled. So the
massive Dirac fermions alone do not contribute a half-quantized Hall conductance
to the system. The system has a minimal longitudinal conductance, and exhibits a
fairly flat plateau of the half-quantized Hall conductance when Fermi level is swiping
the Dirac cone. The plateau is very robust against the disorders. We term the gapless
Dirac cone as parity anomalous semimetal, a semimetal with a half-quantized Hall
conductance. The results suggest a condensed matter realization of the topological
phase with a one-half topological invariant.

\paragraph*{Band structure of a semi-magnetic topological insulator}

A semi-magnetic topological insulator film consists of topological insulator $(\mathrm{Bi,Sb})_{2}\mathrm{Te}_{3}$
and Cr-doped $(\mathrm{Bi,Sb})_{2}\mathrm{Te}_{3}$ grown by molecular-beam epitaxy.
$(\mathrm{Bi,Sb})_{2}\mathrm{Te}_{3}$ is a topological insulator with an energy
gap of about 0.3eV, and hosts a single Dirac cone of the surface electrons \citep{ZhangH-09np,Xia-09np,Chen-09science}.
As shown in Fig. 1(e), the magnetic element Cr was doped on the top surface. The
exchange interaction between the magnetic ion and the surface electrons leads to
nonzero magnetization and opens an energy gap on the top surface electrons \citep{Yu-10science,Chen-10science,Chang-13science,Yoshim-15nc}.
The Fermi energy can be finely tuned by changing the ratio of Bi and Sb such that
it locates within the band gap of the top surface Dirac cone. The material has been
extensively studied since the discovery of topological insulator. The topological
nature of $(\mathrm{Bi,Sb})_{2}\mathrm{Te}_{3}$ can be well described by a tight-binding
model for the electrons of $\mathrm{P}_{z,\uparrow}$ and $\mathrm{P}_{z,\downarrow}$
orbitals from $(\mathrm{Bi,Sb})$ and $\mathrm{Te}$ atoms near the Fermi energy
\citep{ZhangH-09np,Liu2010PRB,Chu-11prb},

\begin{equation}
H_{TI}=\sum_{i}\Psi_{i}^{\dagger}\mathcal{M}\Psi_{i}+\sum_{i,\alpha=x,y,z}\left(\Psi_{i}^{\dagger}\mathcal{T}_{\alpha}\Psi_{i+\alpha}+\Psi_{i+\alpha}^{\dagger}\mathcal{T}_{\alpha}^{\dagger}\Psi_{i}\right)\label{eq:tight-binding-model}
\end{equation}
where $\mathcal{M}=(m_{0}-2\sum_{\alpha}t_{\alpha})\sigma_{0}\tau_{z}$, $\mathcal{T}_{\alpha}=t_{\alpha}\sigma_{0}\tau_{z}-i\frac{\lambda_{\alpha}}{2}\sigma_{\alpha}\tau_{x}$,
$\Psi_{i}^{\dagger}$ and $\Psi_{i}$ are the four-component creation and annihilation
operators at position $i$. The Pauli matrices $\sigma_{\alpha}$ and $\tau_{\alpha}$
act on the spin and orbital indices, respectively. All bands are doubly degenerate
due to the coexistence of both time-reversal and inversion symmetries in the absence
of magnetic doping. It can produce the linear dispersion of the surface states near
the $\Gamma$ point in an open boundary condition. The exchange interaction caused
by Cr doping is given by a $V_{exch}=\sum_{i}\Psi_{i}^{\dagger}V(i)\sigma_{z}\tau_{0}\Psi_{i}$
which is only present on the lattice sites of the top layers with the magnitude as
$V_{z}$. In experiments, the thickness of the Cr-doped $(\mathrm{Bi,Sb})_{2}\mathrm{Te}_{3}$
is about 2nm and $(\mathrm{Bi,Sb})_{2}\mathrm{Te}_{3}$ layer is about 8nm \citep{Mogi-21np}.
We take the periodic boundary condition in the x- and y-direction and open boundary
condition in the z direction, the calculated dispersions are presented in Fig. 2a.
It is noted that there exist a gapless Dirac cone and a gapped Dirac cone within
the bulk gap. The dispersions for the gapless Dirac cone cross at the $\Gamma$ point
and are linear in $k$ around the crossing point. The gapped Dirac cone opens an
energy gap of about $2V_{z}$, which is caused by the exchange interaction of Cr-doping
on the top surface. Numerical caluculation shows that the gapless and gapped states
within the bulk band gap are mainly located on the bottom and top surfaces, respectively.
We check the energy separation between the two bands along the high symmetric lines,
and find that the gapless Dirac cone and gapped Dirac cone are well separated. The
dip at $k_{c}$ indicates that there exists a band mixture. Finite thickness of the
film may cause a tiny gap at the $\Gamma$ point, which decays exponentially in the
thickness approximately\citep{Lu-10prb,Linder-09prb}. With increasing exchange interaction,
the gap is quickly suppressed by several order of magnitudes to negligibly small
(about $10^{-10}$eV. for a thickness $L_{z}=10$nm, see Fig. 2c). It will be smeared
out easily by temperature broadening ($1K$ is about $0.086meV$) in experimental
measurements.

\begin{figure}
\includegraphics[width=8.5cm]{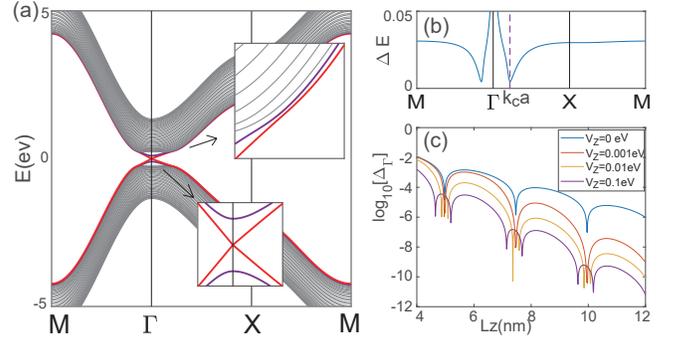}

\caption{The band structure of a semi-magnetic topological insulator. (a) The dispersions
of well separated gapless (red line) and gapped (violet line) Dirac cone in a topological
insulator thin film of 8-nonmagnetic layers plus 2 magnetic-layer. Inset 1): the
dispersions near the $\Gamma$ point; Inset 2): the anti-band crossing point between
the gapless and gapped Dirac cones. (b) The energy separation between the gapless
and gapped Dirac cones. (c) The energy difference of the gapless Dirac cone at the
$\Gamma$ point as a function of the thickness $L_{z}$ for several values of exchange
interaction $V_{z}$ on the top layer. Model parameters: $\lambda_{x}=\lambda_{y}=\lambda_{\parallel}=0.41$
eV, $\lambda_{z}=\lambda_{\perp}=0.44$ eV, and $t_{x}=t_{y}=t_{\parallel}=0.566$
eV, $t_{z}=t_{\perp}=0.40$ eV and $m_{0}=0.28$ eV. $V_{z}=0.1$ eV if with no specific
indication.}
\end{figure}

\paragraph*{The Hall conductance and the Berry curvature}

Using the Kubo formula for the electric conductivity \citep{Mahan-book}, we calculate
the Hall conductance as a function of the chemical potential $\mu_{F}$ numerically
for the tight-binding model in Eq. \ref{eq:tight-binding-model} with a thickness
$L_{z}=10$nm. A fairly flat plateau of $-\frac{e^{2}}{2h}$ appears within the band
gap as shown in Fig. 3a. To figure out the origin of the conductance plateau, we
first note that there exists a full band gap between four lowest energy bands and
the rest at all $k$. These four bands form well-separated band-subspaces and the
Hall conductance as a function of $\mu_{F}$ can be calculated for each band. We
then only focus on the gapless and gapped Dirac cones denoted by red and violet lines
in Fig. 2a. For the gapped Dirac cone, we have a nonzero Hall conductance as $\mu_{F}$
varies, which has its maximal about $0.4\frac{e^{2}}{h}$, but decays to zero quickly
when the band is fully occupied. The maximal value may increase for a thicker film,
but always be lower than $0.5\frac{e^{2}}{h}$. This is consistent with the fact
that the Chern number of a well-defined band in a finite Brillouin zone is always
an integer (including zero) \citep{shen2012topological,Xiao-10rmp}. For the gapless
Dirac cone, the Hall conductance becomes $-0.5\frac{e^{2}}{h}$ within the bulk band
gap, which is larger than the gap of the gapped Dirac cone. Thus the total Hall conductance
within the bulk band gap are mainly contributed by these two bands and the Hall conductance
plateau is attributed to the gapless Dirac cone instead of the gapped Dirac cone.

To explore the topological nature of the gapless Dirac cone and its relation to the
Hall conductance, we studied the Berry curvature of the gapless bands. In the Bloch
states $\left|u_{n,\mathbf{k}}\right\rangle $, the Berry connection and the Berry
curvature are defined as $\mathcal{A}_{n,\alpha}(\mathbf{k})=i\langle u_{n,\mathbf{k}}|\partial_{k_{\alpha}}u_{n,\mathbf{k}}\rangle$
and $\Omega_{z}^{n}(\mathbf{k})=\partial_{k_{x}}\mathcal{A}_{n,y}(\mathbf{k})-\partial_{k_{y}}\mathcal{A}_{n,x}(\mathbf{k})$,
respectively \citep{Xiao-10rmp}. For the gapless Dirac cone, it is found that the
Berry curvature $\Omega_{z}^{n}(\mathbf{k})=0$ within the regime of $k<k_{c}$.
Beyond the regime, it becomes negative and finally vanishes for a larger $k$. Combining
with the band structure, the nonzero Berry curvature mainly originates from hybridization
of the states from the top and bottom layers. The conductance plateau appears when
the chemical potential is located in the regime where the Berry curvature vanishes.

\begin{figure}
\includegraphics[width=8.5cm]{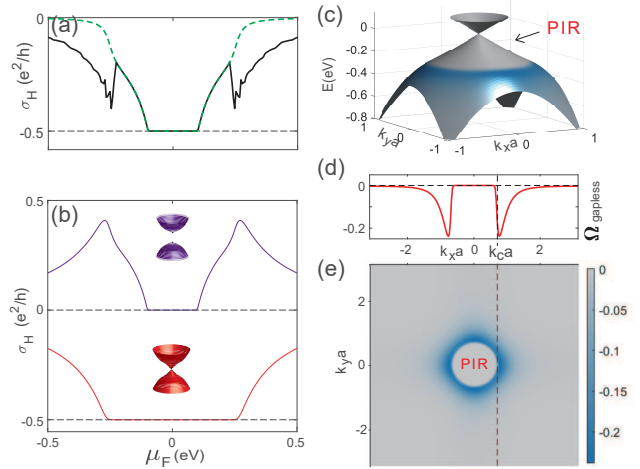}\caption{(a) The Hall conductance as a function of the chemical potential $\mu_{F}$ for the
thin film. The green dashed line is the combination of the Hall conductance of the
gapless and gapped Dirac cones. (b) The Hall conductance of the gapless and gapped
Dirac cones. (c) The gapless Dirac cone in a two-dimensional Brillouin zone. The
color indicates the value of the Berry curvature of the states. (d) The Berry curvature
of the gapless valence band as a function of $k_{x}$ ($k_{y}=0$). (e) The Berry
curvature and the parity-invariant regime (PIR) in the Brillouin zone.}
\end{figure}

\paragraph*{On the parity-invariant regime}

Now we come to discuss the origin of the parity-invariant regime based on the tight-banding
model in Eq. (1). For each wave vector $\mathbf{k}$, the Hamiltonian can be divided
into two parts, $H_{TI}(\mathbf{k})=H_{1d}(\mathbf{k})+H_{S}(\mathbf{k)}$. $H_{1d}$
is equivalent to a one-dimensional lattice model with a gap $m_{0}(k)=m_{0}-4t_{\parallel}\left(\sin^{2}\frac{k_{x}a}{2}+\sin^{2}\frac{k_{y}a}{2}\right)$
\citep{shen2012topological},
\begin{equation}
H_{1d}(\mathbf{k})=\sum_{i_{z}}\left(\Psi_{i_{z},\mathbf{k}}^{\dagger}\mathcal{M}(\mathbf{k})\Psi_{i_{z},\mathbf{k}}+\Psi_{i_{z},\mathbf{k}}^{\dagger}\mathcal{T}_{z}\Psi_{i_{z}+1,\mathbf{k}}+h.c.\right)
\end{equation}
with $\mathcal{M}(\mathbf{k})=\left[m_{0}(k)-2t_{\perp}\right]\sigma_{0}\tau_{z}$,
and $H_{S}(\mathbf{k})=\lambda_{\parallel}\sum_{i_{z},\alpha=x,y}\Psi_{i_{z},\mathbf{k}}^{\dagger}\sin\left(k_{\alpha}a\right)\sigma_{\alpha}\tau_{x}\Psi_{i_{z},\mathbf{k}}$.
For $m_{0}(k)>0$, \textit{i.e.} $k<k_{c}\simeq\sqrt{M/t_{\parallel}}/a$, $H_{1d}$
is topologically non-trivial. There exist a pair of zero energy modes at each side
or near the top surface and bottom surface. Denote $\xi_{s}$ and $\chi_{t}$ the
two eigenvectors of $\sigma_{z}$ and $\tau_{y}$ with eigen values $s=\pm1$ and
$t=\pm1$. The zero energy modes are the eigen vectors $\xi_{s}\otimes\chi_{t}$
of the operator $\sigma_{z}\tau_{y}$ with the eigenvalue $S=st$. The spatial part
of the two states of $S=1$ ($s=t=1$ and $s=t=-1$) are mainly located near the
top surface and the two states of $S=-1$ are located near the top surface, and decay
exponentially to its opposite side \citep{shen2012topological}. By mapping $H_{S}(\mathbf{k})+V_{exch}$
into the the basis of the four states, one obtains the effective Hamiltonian the
gapless Dirac cone $H_{b}(k)=-\lambda_{\parallel}\left(\sin\left(k_{x}a\right)\sigma_{y}-\sin\left(k_{y}a\right)\sigma_{x}\right)$
which is mainly located at the bottom layer and the gapped Dirac cone $H_{t}(k)=+\lambda_{\parallel}\left(\sin\left(k_{x}a\right)\sigma_{y}-\sin\left(k_{y}a\right)\sigma_{x}\right)+V(k)\sigma_{z}$
which is at the top layer. $V(k)$ is given by the expectation of the exchange interaction
$V_{exch}$, and varies with the wave vector, especially when $m_{0}(k)\rightarrow0$
where the wave function of zero energy modes evolve to distribute broadly in space.
In two dimensions, the parity symmetry is defined by $\mathcal{P}H(\mathbf{k})\mathcal{P}^{-1}=H(\hat{M}\mathbf{k})$,
where $\mathcal{P}=i\sigma_{y}$ and $\hat{M}$ is the mirror operator in momentum
space transforming $\mathbf{k}\to\hat{M}\mathbf{k}=(k_{x},-k_{y})$. Thus in the
regime the gapless Dirac cone $\mathcal{P}H_{b}(\mathbf{k})\mathcal{P}^{-1}=H_{b}(\hat{M}\mathbf{k})$
respects the parity symmetry while the gapped Dirac cone $\mathcal{P}H_{t}(\mathbf{k})\mathcal{P}^{-1}\neq H_{t}(\hat{M}\mathbf{k})$
breaks the symmetry due to the presence of $V(k)$. Thus the nontrivial condition
of $m_{0}(k)>0$ defines a parity-invariant regime for the gapless Dirac cone. In
additional to the local parity symmetry, $H_{b}$ also respects a space-time operator
$I_{ST}=C_{2z}T$ where $C_{2z}$ is a twofold rotation about the $z$ axis, and
$T$ is local time reversal operator. $I_{ST}^{2}=+1$ imposes a further constraint
on the Berry curvature, leading to $\Omega_{z}^{n}(\mathbf{k})=0$. When $m_{0}(k)<0$,
$H_{1d}$ becomes topologically trivial. The zero energy modes evolve into the bulk
states with nonzero energy.. The states are no longer the eigen-vectors of the operator
$\sigma_{z}\tau_{y}$ , and break the parity symmetry. Based on this picture, we
propose an effective four-band model

\begin{equation}
H_{F}=\lambda_{||}(\sin k_{x}a\sigma_{y}-\sin k_{y}a\sigma_{x})\gamma_{z}+V(k)\sigma_{z}(\gamma_{z}+1)+H_{\Delta}^{\textrm{eff}}
\end{equation}
where $H_{\Delta}^{\textrm{eff}}=f(k)m_{0}(k)\gamma_{x}$ describes the hybridization
of the states. A Fermi-Dirac-distribution-like factor or the sigmoid function $f(k)=\left[\exp\left(\frac{m_{0}(k)}{T^{*}}\right)+1\right]^{-1}$
is introduced to describe the procedure that the coupling is eventually turned on
at higher energy. $\gamma_{x,z}$ are the Pauli matrices representing the top and
bottom layers. A small $T^{*}$ is a model-specific parameter. The calculated results
show that the model can reproduce the key features of the band structure. The Hall
conductance as a function of $\mu_{F}$ is given by $\sigma_{H}^{S}=\frac{e^{2}}{2h}(S-\cos\phi_{S}(k_{F}^{S}))$.
The part of $S$ is mainly attributed to the hybridization term $H_{\Delta}^{eff}$
and the band splitting $V(k)$. For the gapless band of $S=-1$, $\cos\phi_{S}(k_{F}^{S})=0$
in the parity-invariant regime and $\sigma_{H}^{-1}=-\frac{e^{2}}{2h}$. For the
gapped band of $S=+1$, $\cos\phi_{S}=1$ for the full occupancy and $\sigma_{H}^{+1}=0$.
They are in a good agreement with the numerical results in Fig. 3b. The details are
referred to Ref. \citep{note-SI}.

\paragraph*{Half-quantization and parity symmetry}

Now we present a relation between the half-quantization of the Hall conductance and
the parity symmetry. In the Haldane model and the Wilson fermions, it was found that
the Hall conductance is half-quantized when the chemical potential is located at
the energy crossing point \citep{Haldane1988Model,FuB-22xxx}. Here we prove that
the Hall conductance is one half of an integer if the parity symmetry is respected
near the Fermi level $\mu_{F}$. In two dimensions, the intrinsic Hall conductance
for band $n$ can be expressed in terms of the Berry gauge field in momentum space
\citep{Xiao-10rmp}, 
\begin{equation}
\sigma_{H}^{n}(\mu_{F})=\frac{e^{2}}{h}\int\frac{d^{2}k}{2\pi}\epsilon^{ij}\partial_{k_{i}}\mathcal{A}_{n,j}\theta(\varepsilon_{n,\mathbf{k}}-\mu_{F})
\end{equation}
where $\theta(x)$ is the step function, \textit{i.e.}, $\theta(x)=1$ for $x\leq0$
and zero otherwise. The whole Brillouin zone can be divided into two patches by the
Fermi level $\mu_{F}$, the occupied and unoccupied regime, respectively. By using
the Stokes' theorem, the integral of the Berry curvature over the occupied regime
can be converted into the integrals over the Fermi surface (FS), $\sigma_{H}^{n}(\mu_{F})=\frac{e^{2}}{h}\frac{1}{2\pi}\oint_{FS}d\vec{l}_{F}\cdot\boldsymbol{\mathcal{A}}_{n},$
where $\vec{l}_{F}$ is the wave vector along the Fermi surface. Consider the case
that the Hamiltonian $H(\mathbf{k})$ is invariant under the parity symmetry along
the Fermi surface, \textit{i.e.} $\mathcal{P}H(\mathbf{k})\mathcal{P}^{-1}=H(\hat{M}\mathbf{k})$.
The eigenstates of $H(\mathbf{k})$ at $\mathbf{k}$ and $\hat{M}\mathbf{k}$ on
the Fermi edge must be related by a gauge transformation: $\mathcal{P}|u_{n}(\mathbf{k})\rangle=\exp[i\theta_{n}(\mathbf{k})]|u_{n}(\hat{M}\mathbf{k})\rangle$.
Accordingly, the Berry connection is transformed to $\mathcal{A}_{n,i}(\mathbf{k})=\partial_{k_{i}}\theta_{n}(\mathbf{k})+\sum_{j}J_{ij}\mathcal{A}_{n,j}(\hat{M}\mathbf{k})$
with $J_{ab}=\partial(\hat{M}\mathbf{k})_{b}/\partial k_{a}$. The determinant of
the Jacobian matrix $J_{ab}$ equals $-1$. It follows that $2\sigma_{H}^{n}=\frac{e^{2}}{h}\frac{1}{2\pi}\oint_{FS}d\vec{l}_{F}\cdot\nabla_{\vec{l}_{F}}\theta_{n}(\mathbf{k})=\frac{e^{2}}{h}\nu$.
For a linear dispersion near the Dirac point, the Berry phase around the Fermi surface
is $+\pi$ or $-\pi$, depending on the helicity of the Dirac fermions. Correspondingly
$\nu$ is equal to $+1$ or $-1$. Thus the Hall conductance $\sigma_{H}=\frac{\nu}{2}\frac{e^{2}}{h}$
is half quantized in the case that the parity symmetry is restored near the Fermi
level.

\paragraph*{Disorders and robustness of the quantized Hall conductance}

The robustness of the half-quantized Hall conductance comes from two aspects. One
is the local parity invariance for the gapless Dirac cone. The exchange interaction
is only present at the top surface. The low-energy dispersion of the gapless Dirac
cone is mainly located at the bottom layer, and is less affected by the exchange
interaction, although the part of higher energy is modified. The other aspect is
that the presence of the surface states is mainly attributed to the three-dimensional
band structure of the topological insulator. It is known that the topology of the
band structure of three-dimensional topological insulator is very robust against
the disorders. If the strength of disorders is not strong enough to induce a topological
phase transition, the gapless surface states are still present. In this way the Dirac
point is very stable against the disorders before the phase transition occurs.

To illustrate the robustness of the single Dirac cone in this quasi-two-dimensional
system, we calculated the Hall conductance as a function of the strength of disorders.
We follow the common practice in the study of Anderson localization to introduce
disorder through random non-magnetic on-site energy with a uniform distribution with
$[-W/2,+W/2]$. We calculate the Hall conductance of a disordered square of size
$L_{x}\times L_{y}\times L_{z}$ for a fixed thickness $L_{z}$ by means of the noncommutative
Kubo formula \citep{Prodan-11JPA}
\begin{equation}
\sigma_{H}=i2\pi\mathrm{Tr}\left\{ \mathbf{P}\left[-i[\mathbf{x},\mathbf{P}],-i\left[\mathbf{y},\mathbf{P}\right]\right]\right\} \frac{e^{2}}{h}
\end{equation}
with the periodic boundary conditions along the x and y directions. Here, $\mathbf{x}$
and $\mathbf{y}$ are the coordinate operators and $\mathrm{Tr}\{\cdots\}$ is the
trace over the occupied bands. $\mathbf{P}$ is the projector onto the occupied states
of the system. Figure 4 shows the numerically calculated disorder-averaged Hall conductances
as functions of the disorder strength. In the clean limit, the system exhibits the
half-quantized Hall conductance as expected. With increasing disorder strength, the
Hall conductance remains about $0.5\frac{e^{2}}{h}$ until the disorder strength
$W$ exceeds about 0.83eV. The critical disorder strength is much larger than the
exchange interaction and also larger than the bulk energy gap of topological insulator.
Further increasing the disorder strength, the conductance drops quickly and the system
is expected to be localized by disorders. Thus, the half-quantized Hall conductance
is robust against the disorder. This demonstrates explicitly that the gapless Dirac
cone is quite stable against the disorders.

The presence of impurities will cause the scattering between electron wave functions
which leads to an energy level repulsion effect\citep{Beenakker1997random}. The
stability of the Dirac point is thus equivalent to examine the relative energy level
repulsion between the two states at Dirac point from all other surface and bulk states.
The scatterings between two bottom surface states will not introduce Dirac mass renormalization
due to the presence of the local time reversal symmetry. The scatterings from the
top surface states can also be neglected due to the fact that the two opposite surface
states have exponentially small overlap. The magnetic doping on the top surface will
cause an energy splitting $\sim V_{z}L_{z}^{\mathrm{mag}}/L_{z}$ for the two degenerate
bulk states with $L_{z}^{\mathrm{mag}}$ as the thickness of the magnetic layers.
For sufficiently small $L_{z}^{\mathrm{mag}}/L_{z}$, the energy level repulsion
effect from two nearly degenerate bulk states will be cancelled out. This picture
is verified by a self-energy calculation in the self-consistent Born approximation
based on the effective four-band model \citep{note-SI}. Compared with the two-band
model\citep{Groth2009theory}, the contributions to the Dirac mass renormalization
from the gapped and gapless bands in the four-band model alternate in sign and cancel
each other out at high energy. As a consequence, the Dirac point is still stable
even in the presence of disorder due to the local time-reversal symmetry, and the
accompanying gapped bands.

\begin{figure}
\includegraphics[width=7.5cm]{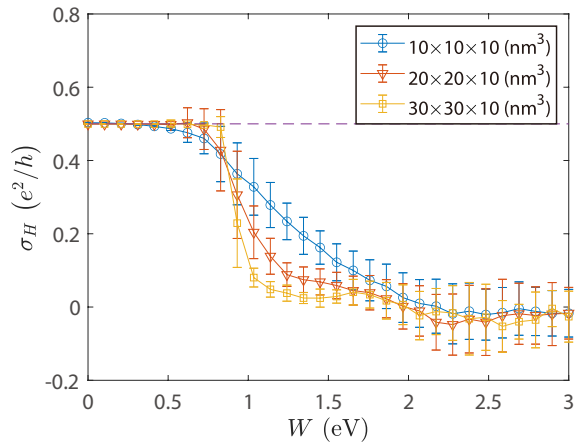}\caption{The Hall conductance as a function of the strength of ) nonmagnetic disorders $U_{i}\sigma_{0}\tau_{0}$
for several lattice sizes $L_{x}\times L_{y}\times L_{z}$ for a fixed $L_{z}=10$nm.
The Fermi level $\mu_{F}=0.01$ eV and the lattice spacing $a=1$nm. 200 random configurations
are adopted to average for each value.}
\end{figure}

\begin{acknowledgments}
The authors would like to thank Junren Shi and Qian Niu for helpful discussions.
The numerical calculations were supported by Center for Computational Science and
Engineering of SUSTech. This work was supported by the Research Grants Council, University
Grants Committee, Hong Kong under Grant Nos. C7012-21G and 17301220 and the National
Key R\&D Program of China under Grant No. 2019YFA0308603.
\end{acknowledgments}

\end{document}